\documentclass[draftclsnofoot, onecolumn, comsoc, 12pt]{IEEEtran}
\usepackage[T1]{fontenc}
\usepackage[utf8]{inputenc}
\usepackage{graphicx,url,cite}
\usepackage[export]{adjustbox} 
\usepackage{placeins} 
\usepackage{amsmath,amssymb,amsfonts,amsthm,bm}
\usepackage{booktabs,multirow,array,stfloats}
\usepackage{algorithm,algpseudocode}
\usepackage{siunitx,xcolor}
\usepackage[hidelinks]{hyperref}
\usepackage{float}
\theoremstyle{remark}

\newcommand{\StarFigTwoWidth}{1\linewidth}
\newcommand{\StarFigThreeWidth}{0.72\linewidth}
\newcommand{\StarFigFourWidth}{1\linewidth}
\newcommand{\StarFigFiveWidth}{1\linewidth}
\newcommand{\StarSingleHeight}{0.72\textheight}
\newcommand{\StarWideHeight}{0.43\textheight}
\title{StarRTT: An End-to-End Real-Time Framework for Blind Zone-Free RTT Prediction in Starlink LEO Satellite Networks}
\author{Jaeyong Lee, Seongjin Jeong, Yusang Park, \\Hyunho Hwang, Sangmin Han, and Wonjae Shin%
\thanks{Jaeyong Lee, Seongjin Jeong, Yusang Park, Hyunho Hwang, Sangmin Han, and Wonjae Shin are with the School of Electrical Engineering, Korea University, Seoul 02841, South Korea (e-mails: jaeyong712@korea.ac.kr; jsj3020@korea.ac.kr; ysparka@korea.ac.kr; hhhwang0503@korea.ac.kr; smhan22@korea.ac.kr; wjshin@korea.ac.kr).}%
\thanks{\emph{Jaeyong Lee and Seongjin Jeong are co-first authors.}}%
}
\begin{document}
\markboth{}{}
\maketitle
\begin{abstract}
Low Earth Orbit (LEO) satellite mega-constellations such as Starlink require frequent handovers due to high orbital velocities, resulting in substantial Round-Trip Time (RTT) fluctuations. Accurate RTT prediction can mitigate these fluctuations, yet it fundamentally relies on continuously updated serving-satellite information. Existing serving-satellite identification methods typically complete identification only at the end of each scheduling cycle, creating an identification blind zone that limits real-time prediction. To overcome this limitation, we develop StarRTT, an end-to-end real-time framework that eliminates the blind zone by integrating per-second serving-satellite identification with RTT prediction. Within this framework, we propose Real-Time Long Short-Term Memory (RT-LSTM), an LSTM-based RTT prediction model designed to capture both non-stationary RTT evolution and time-varying geometric relationships among the serving satellite, user terminal, and gateways. Specifically, a historical LSTM encodes historical RTT observations and real-time serving-satellite and gateway-relative features, while a future LSTM processes Two-Line Element (TLE)-propagated future features of the identified serving satellite up to the next scheduling instant.  Under strict real-time back-testing, RT-LSTM achieves a Mean Absolute Error (MAE) of \SI{1.405}{ms} averaged over all \num{15} prediction horizons and \num{15} reference time-slot indices, reducing the error by \SI{4.3}{\%}, \SI{11.6}{\%}, and \SI{22.1}{\%} compared with the existing baseline models, LSTM, StarNet, and Attention-based Spatiotemporal Model (ASTM), respectively. Notably, the reduction over LSTM reaches \SI{12.5}{\%} at the scheduling instant, highlighting the benefit of real-time serving-satellite information when recent RTT observations are limited.
\end{abstract}


\section{Introduction}
The rapid deployment of Low Earth Orbit (LEO) mega-constellations, such as SpaceX's Starlink, is enabling low-latency broadband Internet services on a global scale \cite{starlink_web}. LEO satellites operate at altitudes of approximately $\num{500}$ to $\num{1200}$ km, offering low communication latency due to short propagation distances. However, each satellite covers only a limited geographic area, requiring constellations of thousands of satellites across multiple orbital planes for global connectivity. These satellites also travel at approximately \SI{7.5}{km/s} to maintain their orbits, causing frequent handovers at the ground User Terminal (UT). Specifically, Starlink's scheduling cycle is $\num{15}$ seconds, and at each scheduling instant the system decides whether to perform a handover. These scheduling instants occur at seconds $\num{12}$, $\num{27}$, $\num{42}$, and $\num{57}$ of every minute~\cite{tanveer2023making}. Our empirical measurements further reveal that the average connection duration to a single serving satellite is approximately $\num{28.5}$ seconds. Consequently, frequent handovers are a major source of latency instability in LEO Satellite Networks (LSNs).

\begin{figure}[!t]
\centering
\includegraphics[width=0.55\linewidth]{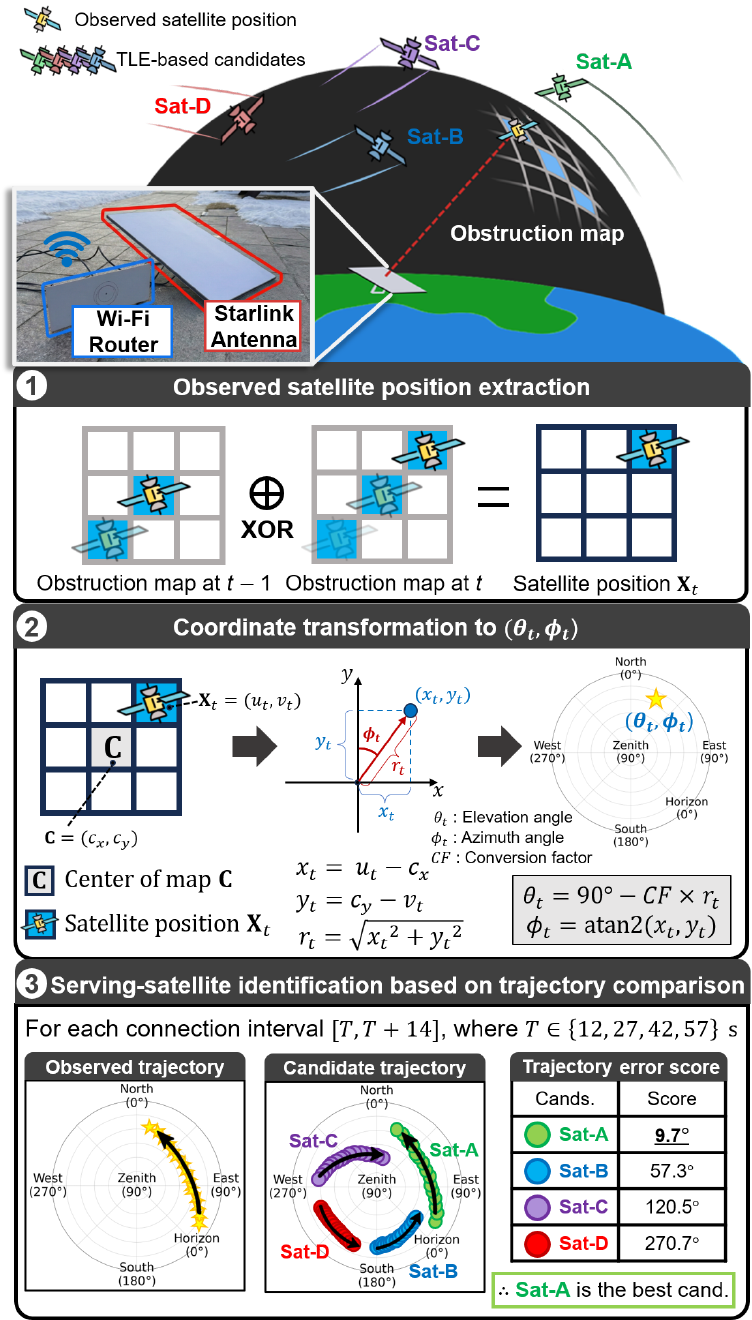}
\caption{Serving-satellite identification using per-second XOR operations on consecutive obstruction maps. The observed position $\mathbf{X}_t$ is converted into $(\theta_t,\phi_t)$, and the observed trajectory is compared with TLE-based candidate trajectories to identify the serving satellite.}
\label{fig:fov}
\end{figure}

A key metric that captures this latency instability is the Round-Trip Time (RTT), defined as the total duration for a data packet to travel from its source to its destination and for the corresponding acknowledgment (ACK) to return. Since each handover changes the serving satellite along the communication path, RTT exhibits abrupt fluctuations in LSNs. Accurate RTT prediction is therefore critical for proactive network management, such as retransmission timer calibration and congestion window adjustment. However, RTT prediction in LSNs poses fundamentally different challenges from terrestrial networks: each handover discontinuously changes the serving satellite, altering both propagation path and distance, thereby making RTT highly non-stationary. Capturing such non-stationary behavior requires identifying which satellite is currently serving the UT. In practice, however, operators do not disclose this information, forcing prediction systems to rely on indirect estimation.
Serving satellites are therefore indirectly identified by matching UT observations with predicted trajectories generated from publicly available Two-Line Element (TLE) data. To date, two representative identification methods have been proposed. The first, proposed by Ahangarpour et al.~\cite{fov}, extracts the serving satellite's trajectory using per-second bitwise Exclusive-OR (XOR) operations on the UT's obstruction map, as shown in Fig.~\ref{fig:fov}, and computes a trajectory error score at the start, mid, and end of each $\num{15}$-second scheduling cycle to identify the satellite with the minimum score. The second, StarNet \cite{starnet}, compares the complete trajectory sequence observed over an entire $\num{15}$-second window against TLE-based candidate trajectories using the Dynamic Time Warping (DTW) algorithm to determine the best matching satellite. However, both methods operate in a \textit{scheduling-cycle-based} manner, creating an \textit{identification blind zone} of up to $\num{14}$ seconds within each $\num{15}$-second scheduling cycle during which no updated serving-satellite information is available.
This structural limitation fundamentally prevents reliable real-time RTT prediction, regardless of predictor accuracy.
\begin{figure}[!t]
\centering
\includegraphics[width=\StarFigTwoWidth,height=\StarSingleHeight,keepaspectratio]{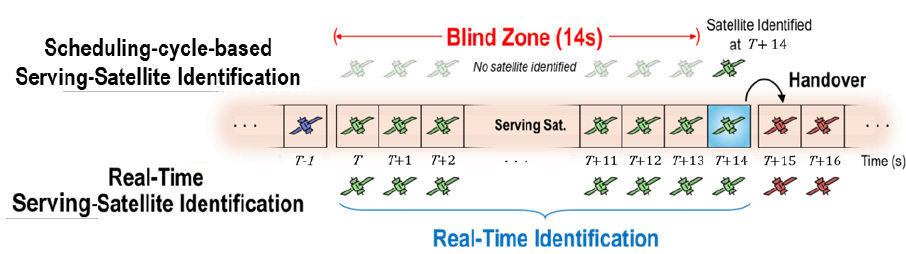}

\caption{Timeline comparison between conventional scheduling-cycle-based serving-satellite identification and the proposed real-time framework. Scheduling-cycle-based methods incur a blind zone lasting $\num{14}$ seconds, whereas the proposed method enables reliable identification for real-time prediction.}

\label{fig:time_table}
\end{figure}
The RTT in LSNs follows complex spatiotemporal patterns driven by the combination of satellite orbital motion and periodic handovers. Among existing methods, the Attention-based Spatiotemporal Model (ASTM) combines a Transformer with Dynamic Graph Convolutional Networks (DGCN) to jointly capture temporal and spatial features, demonstrating the potential of RTT prediction for throughput improvement \cite{astm}. However, ASTM also adopts the serving-satellite identification method of Ahangarpour et al., which relies entirely on scheduling-cycle-based identification. As a result, during the blind zone, this model must rely on stale identification results from the previous cycle.

Even if the serving satellite is accurately identified at the end of a scheduling cycle, the result may not remain valid thereafter because the serving satellite can change. In this case, the prediction model cannot utilize current serving-satellite information. Therefore, improving identification accuracy alone is insufficient; the temporal resolution of the identification stage fundamentally limits the real-time prediction capability of the overall system.

To overcome this limitation, this paper develops a novel real-time framework, termed StarRTT, that jointly performs per-second serving-satellite identification and multi-step RTT prediction, thereby eliminating the blind zone inherent in scheduling-cycle-based approaches. Within this framework, we propose Real-Time Long Short-Term Memory (RT-LSTM) as the RTT prediction model~\cite{lstm}. As illustrated in Fig.~\ref{fig:time_table}, the proposed framework updates the serving-satellite identification result every second, whereas existing scheduling-cycle-based methods can identify the serving satellite only at the final second of each $\num{15}$-second scheduling cycle. RT-LSTM predicts RTT over the next $\num{15}$ seconds every second using historical RTT observations, real-time serving-satellite and gateway-relative features, and TLE-propagated future features of the identified serving satellite up to the next scheduling instant. The prediction is performed at reference time $t$, and the corresponding reference time-slot index is defined as $t \bmod 15$. The main contributions of this paper are summarized as follows.

\begin{itemize}
\setlength{\topsep}{1pt}
\setlength{\itemsep}{1pt}
\setlength{\parsep}{0pt}
\setlength{\parskip}{0pt}
\item \textbf{End-to-End Real-Time RTT Prediction Framework:} We develop StarRTT, a unified framework that integrates per-second serving-satellite identification and multi-step RTT prediction into a single real-time pipeline, eliminating the identification blind zone and enabling continuous real-time RTT prediction throughout each scheduling cycle.

\item \textbf{Real-Time Serving-Satellite Identification and TLE-Propagated Future Feature Fusion:} A per-second serving-satellite identification algorithm is designed to eliminate the blind zone, achieving a minimum agreement of \SI{98.5}{\%} with the final identification result obtained from the full $\num{15}$-second trajectory of each scheduling cycle. RT-LSTM combines historical RTT observations and real-time serving-satellite features with TLE-propagated future features through dual LSTM branches and gated fusion, enabling per-second RTT prediction over the next 15 seconds.

\item \textbf{Gateway-Relative Feature Encoding and Performance Validation:} Multi-generation analysis across Starlink V1.0, V1.5, and V2.0-mini reveals distinct RTT characteristics within and outside an empirically identified bent-pipe coverage area. Motivated by this observation, RT-LSTM incorporates gateway-relative features. Under strict real-time back-testing, RT-LSTM achieves a Mean Absolute Error (MAE) of \SI{1.405}{ms} averaged across all $\num{15}$ prediction horizons and $\num{15}$ reference time-slot indices, reducing the MAE by \SI{4.3}{\%}, \SI{11.6}{\%}, and \SI{22.1}{\%} compared with the existing baseline models, LSTM, StarNet, and ASTM, respectively.
\end{itemize}

\section{Proposed Real-Time Serving-Satellite Identification and RTT Prediction}
\label{sec:methodology}

\subsection{Proposed Serving-Satellite Identification Method}
\label{sec:identifying_method}

In this paper, we propose a real-time serving-satellite identification algorithm. Unlike the conventional scheduling-cycle-based method, which calculates the trajectory error score using only three discrete observations (start, mid, and end points) \cite{fov}, the proposed framework updates the serving-satellite identification every second, thereby enabling real-time RTT prediction. To achieve this, we develop a refined scoring mechanism based on a temporal importance analysis of two constituent metrics: angular distance and directional difference between the observed satellite trajectory and the TLE-propagated trajectory \cite{fov}. The observed coordinates $(\theta_{\mathsf{obs}}, \phi_{\mathsf{obs}})$ are obtained using the procedure shown in Fig.~\ref{fig:fov}. The angular distance $\delta$ between these coordinates and the TLE-propagated coordinates $(\theta_{\mathsf{sat}}, \phi_{\mathsf{sat}})$ is defined as:
\begin{equation}
\delta \!=\! \arccos \left( \sin \theta_{\mathsf{obs}} \sin \theta_{\mathsf{sat}}\!+\! \cos \theta_{\mathsf{obs}} \cos \theta_{\mathsf{sat}} \cos(\phi_{\mathsf{obs}} \!-\! \phi_{\mathsf{sat}}) \right).
\end{equation}
Furthermore, the directional difference is defined as the minimum circular angle separation $\Delta\beta$ between the observed trajectory direction $\beta_{\mathsf{obs}}$ and the TLE-based trajectory direction $\beta_{\mathsf{sat}}$ \cite{fov-github}:
\begin{equation}
\Delta\beta = \min(|\beta_{\mathsf{obs}} - \beta_{\mathsf{sat}}|, \ang{360} - |\beta_{\mathsf{obs}} - \beta_{\mathsf{sat}}|).
\label{eq:direction_difference}
\end{equation}

\begin{algorithm}[t]
\caption{Real-time Serving-Satellite Identification}\label{alg:satellite_identification}

\begin{algorithmic}[1]
\Statex \textbf{Input:} Observed trajectory \\$\mathcal{T}_{\mathsf{obs}} = \{(\theta_{\mathsf{obs},i}, \phi_{\mathsf{obs},i}, t_i)\}_{i=0}^{N}$ and candidate satellite set $\mathcal{S}_{\mathsf{tle}}$ from TLE data
\Statex \textbf{Output:} Best identified serving satellite ${s}_{\mathsf{best}}$

\State \textbf{Initialize:} $\mathcal{C}_{\mathsf{cand}} \leftarrow \emptyset$

\For{\textbf{each} satellite $s \in \mathcal{S}_{\mathsf{tle}}$}
\State Compute TLE-propagated trajectory $\mathcal{T}_{\mathsf{sat}} = \{(\theta_{\mathsf{sat},i}, \phi_{\mathsf{sat},i})\}_{i=0}^{N}$ at $t_0, \dots , t_N$
\State Calculate angular distances $\delta_i$ between $\mathcal{T}_{\mathsf{obs}}[i]$ and $\mathcal{T}_{\mathsf{sat}}[i]$ using (1)
\State Initialize score: $\mathsf{Score}(s) \leftarrow \sum_{i=0}^{N} \delta_i$

\If{$N \ge 5$}
\State Calculate directional difference $\Delta\beta$ between $\mathcal{T}_{\mathsf{obs}}$ and $\mathcal{T}_{\mathsf{sat}}$ using (2)
\State $\mathsf{Score}(s) \leftarrow \mathsf{Score}(s) + \Delta\beta$
\EndIf

\State $\mathcal{C}_{\mathsf{cand}} \leftarrow \mathcal{C}_{\mathsf{cand}} \cup \{s\}$
\EndFor

\State $s_{\mathsf{best}} \leftarrow \operatorname*{arg\,min}\limits_{s \in \mathcal{C}_{\mathsf{cand}}} \,\mathsf{Score}(s)$
\State \Return $s_{\mathsf{best}}$

\end{algorithmic}
\end{algorithm}
The proposed algorithm assumes that no handover occurs after each scheduling instant $T$ (e.g., at $12$, $27$, $42$, and $57$ seconds of each minute) until the next scheduling instant.\footnote{Starlink user terminals also support dynamic beam switching upon serving-link obstruction, allowing switching at arbitrary times unlike regular scheduling-instant-based handovers~\cite{starlink2025beamswitching}.} It relies exclusively on the angular distance from $T$ to $T+4$. Subsequently, from $T+5$ to $T+14$, both angular distance and directional difference are jointly incorporated. This time-dependent metric selection is primarily motivated by the limited pixel resolution of the obstruction map, which makes the trajectory's directional estimates highly susceptible to distortion during short observation intervals. Therefore, during the initial stage of each scheduling cycle, excluding the directional difference and relying solely on the angular distance yields a higher identification accuracy. As additional trajectory points accumulate, however, directional information becomes increasingly reliable, and incorporating directional difference significantly improves overall identification accuracy. A summarized overview of the proposed identification and scoring procedure is presented in Algorithm~\ref{alg:satellite_identification}.

\begin{figure}[!t]
\centering
\includegraphics[width=\StarFigThreeWidth,height=\StarSingleHeight,keepaspectratio,trim=0cm 0.04cm 0cm 0cm,clip]{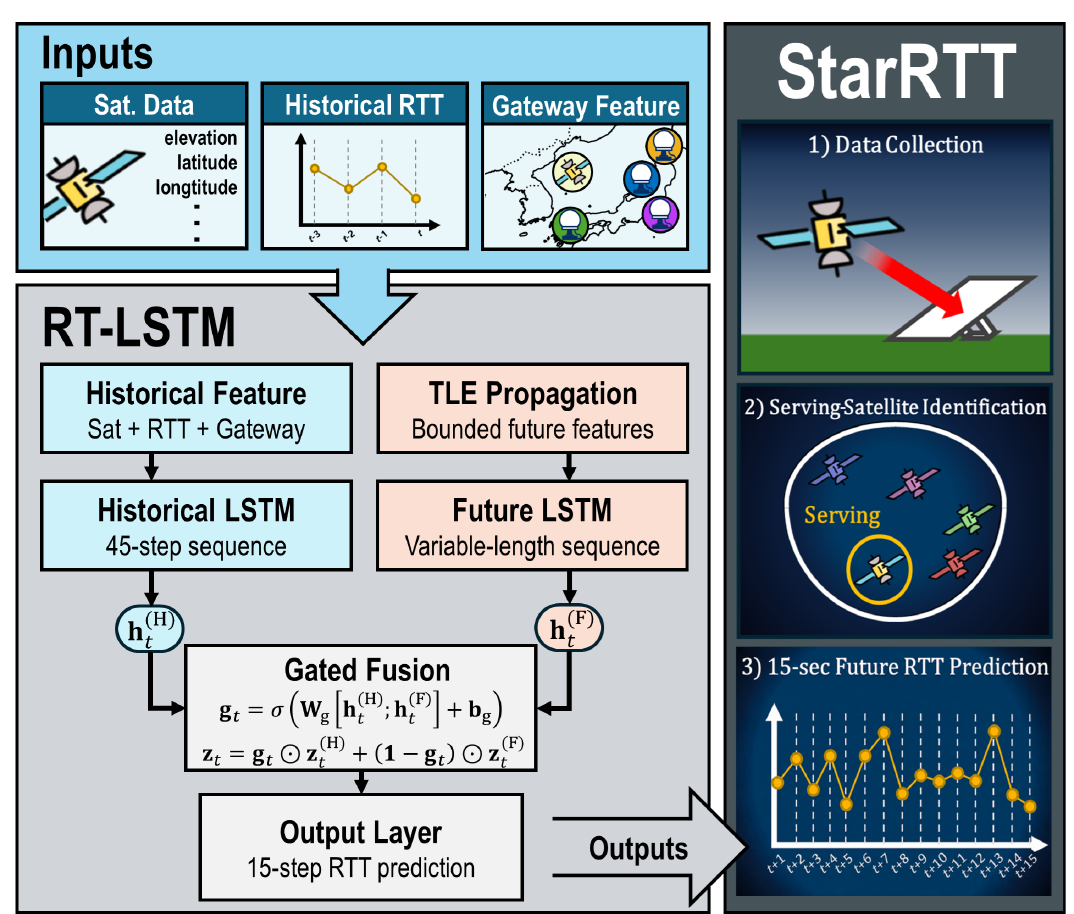}
\caption{System architecture of the StarRTT framework, illustrating real-time serving-satellite identification and the proposed RT-LSTM with dual LSTM branches for encoding historical RTT observations, serving-satellite, and gateway-relative features, and TLE-propagated future features, followed by gated fusion for RTT prediction.}
\label{fig:flow_chart}
\end{figure}

\subsection{RT-LSTM Prediction Architecture}

As shown in Fig.~\ref{fig:flow_chart}, the proposed RT-LSTM employs historical and future LSTM branches, whose representations are combined through gated fusion for multi-step RTT prediction over a $\num{15}$-second horizon.

\subsubsection{Real-time serving-satellite and gateway-relative feature encoding}
\begingroup
The features of the serving satellite identified by the proposed real-time serving-satellite identification algorithm are incorporated into the historical input sequence of RT-LSTM. Together with the historical RTT observations, RT-LSTM uses serving-satellite features and gateway-relative features that characterize the geometric relationships among the serving satellite, UT, and gateways. For each reference time $t$, RT-LSTM constructs a historical input sequence using the available observations up to $t$. Formally, at each historical time step $\tau \in [t-44,t]$, the historical feature vector $\mathbf{x}^{(\mathrm{H})}_{\tau}$ is constructed by concatenating the RTT features $\mathbf{r}_{\tau}^{(\mathrm{H})}$, serving-satellite features $\mathbf{s}_{\tau}^{(\mathrm{H})}$, and gateway-relative features $\mathbf{w}_{\tau}^{(\mathrm{H})}$:
\begin{equation}
\mathbf{x}^{(\mathrm{H})}_{\tau} =
\left[
\mathbf{r}_{\tau}^{(\mathrm{H})},
\mathbf{s}_{\tau}^{(\mathrm{H})},
\mathbf{w}_{\tau}^{(\mathrm{H})}
\right].
\end{equation}
The historical window length was set to $\num{45}$ seconds to match the historical context duration used by ASTM~\cite{tian_github}, which consists of $\num{3000}$ time steps at a temporal resolution of \SI{0.015}{s}. The resulting $\num{45}$-step historical input sequence is represented as
\begin{equation}
\mathbf{X}^{(\mathrm{H})}_{t} =
\left[
\mathbf{x}^{(\mathrm{H})}_{t-44},
\ldots,
\mathbf{x}^{(\mathrm{H})}_{t}
\right],
\end{equation}
which is subsequently encoded by the historical LSTM branch. The historical sequence is processed by a two-layer LSTM:
\begin{equation}
\mathbf{h}^{(\mathrm{H})}_{t} =
\operatorname{LSTM}^{(\mathrm{H})}
\left(
\mathbf{X}^{(\mathrm{H})}_{t}
\right),
\end{equation} 
where $\mathbf{h}^{(\mathrm{H})}_{t}$ denotes the final hidden state of the historical LSTM.
\endgroup

\subsubsection{Gateway-relative feature augmentation}
Section III-B indicates that RTT differs between the inferred bent-pipe and Inter-Satellite Link (ISL) modes. We infer the link mode based on whether the serving satellite lies within both the UT's Field of View (FOV) and the feeder-link visibility of a considered gateway. To reflect this distinction, we augment the input feature space with gateway-relative parameters.
Specifically, elevation angle, azimuth angle, and distance to bent-pipe gateways are added to the input sequence. These features provide essential spatial context, enabling the model to implicitly capture RTT disparities associated with link mode.

\subsubsection{Future feature encoding and gated fusion}
Unlike models that rely solely on historical observations, RT-LSTM additionally incorporates future features of the serving satellite identified at the reference time. Specifically, the orbit of the identified serving satellite is propagated using TLE data to construct a variable-length future feature sequence extending from $t+1$ to one second before the next scheduling instant~\cite{celestrak}. At each future time step $\tau$, the future feature vector is constructed as
\begin{equation}
\mathbf{x}^{(\mathrm{F})}_{\tau} =
\left[
\mathbf{s}^{(\mathrm{F})}_{\tau},
\mathbf{w}^{(\mathrm{F})}_{\tau}
\right],
\end{equation}
where $\mathbf{s}^{(\mathrm{F})}_{\tau}$ and $\mathbf{w}^{(\mathrm{F})}_{\tau}$ denote the TLE-propagated serving-satellite features and gateway-relative features, respectively. Let $b(t)$ denote the first scheduling instant after $t$. The future input sequence is defined as
\begin{equation}
\mathbf{X}^{(\mathrm{F})}_{t} =
\left[
\mathbf{x}^{(\mathrm{F})}_{t+1},
\ldots,
\mathbf{x}^{(\mathrm{F})}_{b(t)-1}
\right].
\end{equation}
The future sequence is encoded by the future LSTM as
\begin{equation}
\mathbf{h}^{(\mathrm{F})}_{t} =
\operatorname{LSTM}^{(\mathrm{F})}
\left(
\mathbf{X}^{(\mathrm{F})}_{t}
\right),
\end{equation}
where $\mathbf{h}^{(\mathrm{F})}_{t}$ denotes the final hidden state of the future LSTM.

The historical and future hidden states are projected into a common latent space:
\begin{equation}
\mathbf{z}^{(\mathrm{H})}_{t} =
\mathbf{W}^{(\mathrm{H})}\mathbf{h}^{(\mathrm{H})}_{t} +
\mathbf{b}^{(\mathrm{H})}, \quad
\mathbf{z}^{(\mathrm{F})}_{t} =
\mathbf{W}^{(\mathrm{F})}\mathbf{h}^{(\mathrm{F})}_{t} +
\mathbf{b}^{(\mathrm{F})}.
\end{equation}
Here, $\mathbf{W}^{(\mathrm{H})}$ and $\mathbf{W}^{(\mathrm{F})}$ are learnable projection matrices, while $\mathbf{b}^{(\mathrm{H})}$ and $\mathbf{b}^{(\mathrm{F})}$ are their corresponding bias vectors.

\begingroup
\setlength{\abovedisplayskip}{3pt}
\setlength{\belowdisplayskip}{3pt}
\setlength{\abovedisplayshortskip}{2pt}
\setlength{\belowdisplayshortskip}{2pt}

A feature-wise fusion gate is computed as
\begin{equation}
    \mathbf{g}_{t} = \sigma \left( \mathbf{W}_{g} \left[ \mathbf{h}^{(\mathrm{H})}_{t}; \mathbf{h}^{(\mathrm{F})}_{t} \right] + \mathbf{b}_{g} \right), \label{eq:fusion_gate}
\end{equation}
where $\mathbf{W}_{g}$ and $\mathbf{b}_{g}$ are the learnable weight matrix and bias vector of the fusion gate, respectively. The notation $[\cdot;\cdot]$ denotes vector concatenation, and $\sigma(\cdot)$ denotes the element-wise sigmoid function. The fused representation is obtained as
\begin{equation}
    \mathbf{z}_{t} = \mathbf{g}_{t} \odot \mathbf{z}^{(\mathrm{H})}_{t} + (\mathbf{1} - \mathbf{g}_{t}) \odot \mathbf{z}^{(\mathrm{F})}_{t}, \label{eq:fused_rep}
\end{equation}
where $\odot$ denotes element-wise multiplication and $\mathbf{1}$ denotes an all-ones vector. Accordingly, $\mathbf{g}_{t}$ controls the contribution of the historical representation, while $\mathbf{1} - \mathbf{g}_{t}$ controls that of the future representation. The fused representation is finally mapped to the RTT predictions for the next 15 seconds. The input features are categorized into three groups:
\begin{itemize}
\setlength{\topsep}{1pt}
\setlength{\itemsep}{1pt}
\setlength{\parsep}{0pt}
\setlength{\parskip}{0pt}
\item \textbf{RTT features:} normalized RTT, RTT delta, 3-second rolling mean, and 3-second rolling standard deviation.
\item \textbf{Serving-satellite data features:} serving-satellite position (latitude, longitude, altitude), elevation angle, azimuth angle, distance, and relative velocity to the user terminal (Seoul), time to exit the FOV of the UT, and days since launch.
\item \textbf{Gateway-relative features:} elevation angle, azimuth angle, and distance to four bent-pipe gateways (Yamaguchi, Akita, Otaru, and Hitachinaka).
\end{itemize}
\endgroup
\begin{table}[!t]
\addtocounter{table}{0} 
\centering
\caption{Experimental Environment.}
\label{tab:env}
\resizebox{0.78\linewidth}{!}{%
\begin{tabular}{ll}
\toprule
\textbf{Parameter} & \textbf{Value} \\
\midrule
Collection Period (UTC) & 2026-03-18 05:53 - 2026-03-23 00:20 \\
UT Latitude & \ang{37.5825615} \\
UT Longitude & \ang{127.0262699} \\
UT Altitude & \SI{90.0}{m} \\
Model / Subscription & STANDARD 4X (Gen 3) / Residential \\
Network Load & iPerf3 TCP downlink w/o bandwidth limitation\\
\bottomrule
\end{tabular}%
}
\end{table}
\begin{table}[!t]
\addtocounter{table}{0} 
\centering
\setlength{\tabcolsep}{2pt}
\caption{Comparison of Model Input Timelines.}
\label{tab:timeline_comparison}
\renewcommand{\arraystretch}{1.3}
\resizebox{\textwidth}{!}{%
\begin{tabular}{l c c c c}
\toprule
\multirow{2}{*}{\begin{tabular}{@{}c@{}}\textbf{Model \textbackslash\ Timeline}\end{tabular}} & \textbf{Past windows} & \textbf{Current window before $t$} & \textbf{Reference time} & \textbf{Bounded future} \\
& $[t_{\text{start}}= t-44, b(t)-16]$ & $[b(t)-15, t-1]$ & $[t]$ & $[t+1, b(t)-1]$ \\
\midrule
LSTM \cite{lstm} & \begin{tabular}{@{}c@{}}RTT, Satellite data\end{tabular} & RTT & RTT & $\times$ \\
\midrule
ASTM \cite{astm} & \begin{tabular}{@{}c@{}}RTT, Satellite data, DGCN\end{tabular} & RTT & RTT & $\times$ \\
\midrule
StarNet \cite{starnet} & \begin{tabular}{@{}c@{}}RTT, Satellite data, $t_{\text{start}} = t - 29$\end{tabular} & RTT & RTT & $\times$ \\
\midrule
RT-LSTM (Proposed) & \begin{tabular}{@{}c@{}}RTT, Gateway-relative, \\ Satellite data\end{tabular} & \begin{tabular}{@{}c@{}}RTT, Gateway-relative, \\ Satellite data\end{tabular} & \begin{tabular}{@{}c@{}}RTT, Gateway-relative, \\ Satellite data\end{tabular} & \begin{tabular}{@{}c@{}}Gateway-relative, \\ Satellite data\end{tabular} \\
\bottomrule
\end{tabular}%
}

\end{table}
\section{Empirical Analysis and Experimental Results}

\subsection{Experimental Environment}

The experiments were conducted using the environmental setup summarized in Table~\ref{tab:env}. Specifically, the data collection was performed at Korea University in Seoul, South Korea, yielding approximately $\num{114.5}$ hours of continuous measurements. To ensure strict one-second time alignment across all datasets, timestamps were synchronized with Google's Network Time Protocol (NTP) server every five minutes, rather than relying solely on the local system clock. We utilized SpaceX Starlink gRPC tools to extract the obstruction map and dish telemetry \cite{grpc}. The center of the obstruction map $\mathbf{C}$ and uniform conversion factor (CF) shown in Fig.~\ref{fig:fov} were calibrated to $(\num{61},\num{61})$ and $\ang{110}$ across $\num{85}$ pixels, respectively~\cite{fov}. During the data-collection period, we maintained a continuous TCP downlink from an Amazon Web Services (AWS) cloud server in Tokyo region using a single stream without bandwidth limitation via iPerf3 \cite{iperf3}. Under this controlled network load, RTT was measured specifically from the Starlink Router to the Point of Presence (PoP), while obstruction map pixel coordinates were collected every second using XOR operations.
\begin{figure}[!t]
\centering
\includegraphics[width=\StarFigFourWidth,height=\StarWideHeight,keepaspectratio]{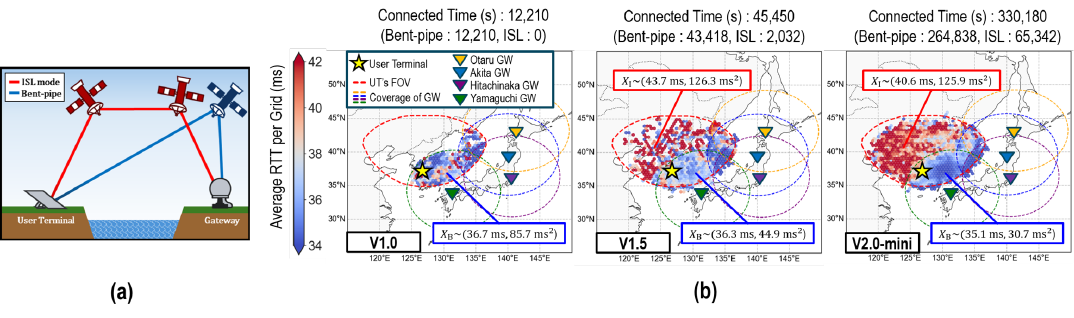}

\caption{Empirical analysis of Starlink link mode and RTT across satellite generations (V1.0, V1.5, V2.0-mini). (a) Architectural comparison of bent-pipe and ISL modes. (b) Spatial distribution of average RTT and connected time, highlighting the bent-pipe coverage area formed by the geometric intersection of the UT's FOV and bent-pipe gateways.}

\label{fig:routing_dynamics}
\end{figure}
This work targets proactive RTT prediction over the next $\num{15}$ seconds, with the prediction updated every second. As summarized in Table~\ref{tab:timeline_comparison}, all models are evaluated under strict conditions that adhere to their real-time data availability constraints. Specifically, scheduling-cycle-based baselines are restricted to satellite information from previous windows and rely only on RTT observations available up to the reference time, whereas the proposed real-time framework seamlessly integrates serving-satellite and gateway-relative features at every second within the current scheduling cycle, including up to one second before the next scheduling instant.

For a consistent comparison, we evaluate the identification results of the three methods using the three-point trajectory error score of Ahangarpour et al.~\cite{fov}. The resulting average scores are $\ang{25.5}$, $\ang{12.7}$, and $\ang{7.1}$ for StarNet~\cite{starnet}, Ahangarpour et al.~\cite{fov}, and the proposed method, where a lower score indicates a closer match between the observed and TLE-propagated trajectories. This lowest score achieved by the proposed method is attributed to the refined calibration of $\mathbf{C}$ and CF. Moreover, at every second within a scheduling cycle, the proposed real-time serving-satellite identification result shows at least \SI{98.5}{\%} agreement with the final identification result obtained from the full $\num{15}$-second trajectory of each scheduling cycle. The high agreement indicates reliable serving-satellite identification without waiting for the full $\num{15}$-second trajectory.

\begin{figure}[!t]
\centering
\includegraphics[width=\StarFigFiveWidth,height=\StarWideHeight,keepaspectratio]{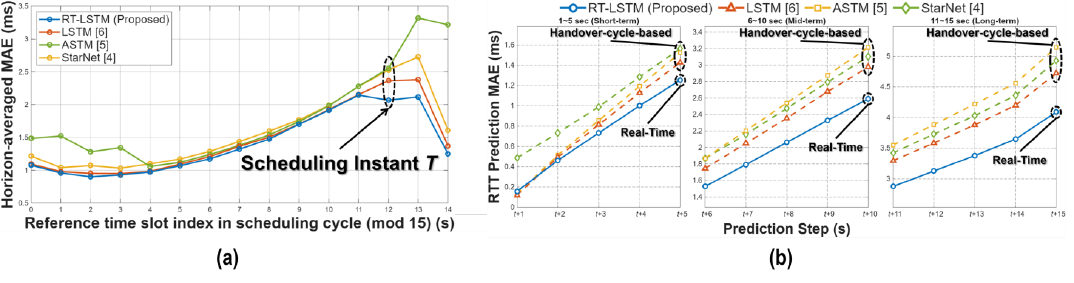}

\caption{Comparison of RTT prediction performance under strict real-time back-testing. (a) Comparison of the MAE, averaged over the $\num{15}$-step prediction horizon, for all evaluated models at each reference time-slot index within a $\num{15}$-second scheduling cycle. (b) Per-step MAE comparison at reference time-slot index $\num{12}$ (the scheduling instant) over the $\num{15}$-step prediction horizon, divided into short-term, mid-term, and long-term intervals.}

\label{fig:mae_combined}
\end{figure}
\subsection{Empirical Analysis of Link Mode and RTT}
\label{sec:routing_analysis}

Since Starlink V1.0 satellites lack ISL hardware, they operate exclusively in bent-pipe mode, whereas both V1.5 and V2.0-mini satellites support ISL \cite{starlink_fov}. As shown in Fig.~\ref{fig:routing_dynamics}, V1.0 satellites establish connections strictly within the UT-Gateway intersection region. Based on the V1.0 observations, and assuming that the same geometric constraints on bent-pipe connectivity apply across satellite generations, we define this region as a bent-pipe coverage area, bounded by a minimum gateway elevation angle of $\ang{35}$ and a minimum UT elevation angle of $\ang{25}$. For V1.5 and V2.0-mini, connections within the UT's FOV but outside this area are classified as inferred ISL cases based on these geometric constraints, yielding bent-pipe-to-ISL ratios of $\num{21}$:$\num{1}$ and $\num{4}$:$\num{1}$, respectively. Because V1.5 and V2.0-mini satellites are uniformly distributed across the same orbital shells, the difference in these inferred link-mode ratios suggests a generation-dependent difference in link utilization that cannot be explained solely by satellite geometry.\footnotemark
Within the geometric intersection of the UT's FOV and the coverage areas of the Japanese gateways (Yamaguchi, Akita, Otaru, and Hitachinaka), the measurements exhibit an average RTT of approximately \SI{35}{ms}. In contrast, samples classified as inferred ISL cases outside this intersection region exhibit an average RTT above \SI{40}{ms}. Thus, the bent-pipe and ISL modes exhibit distinguishable RTT characteristics in the measured data.

\subsection{RTT Prediction Results}
\label{sec:backtest}

Fig.~\ref{fig:mae_combined}(a) compares the four models in terms of MAE, averaged over the $\num{15}$-step prediction horizon, for each of the $\num{15}$ reference time-slot indices under the strict real-time back-testing protocol.\footnotetext{The bent-pipe and ISL modes are inferred primarily from geometric constraints, with the observed RTT distributions serving as supporting evidence. These inferred modes are used only for our analysis and do not represent ground-truth routing decisions within the Starlink network.} Fig.~\ref{fig:mae_combined}(b) further presents the per-step MAE from $t+1$ to $t+15$ at reference times $t$ satisfying $t \bmod 15 = 12$, i.e., reference time-slot index $\num{12}$ corresponding to the scheduling instant.

\subsubsection{Advantage of real-time serving-satellite information}
As shown in Fig.~\ref{fig:mae_combined}(a), RT-LSTM achieves lower horizon-averaged MAE than all baseline models at every reference time-slot index. When averaged across all $\num{15}$ prediction horizons and $\num{15}$ reference time-slot indices, RT-LSTM reduces the MAE by \SI{4.3}{\%}, \SI{11.6}{\%}, and \SI{22.1}{\%} compared with the existing baseline models, LSTM, StarNet, and ASTM, respectively. The improvement over LSTM is particularly pronounced at reference time-slot indices $\num{12}$--$\num{14}$, corresponding to the scheduling instant and the immediately following two seconds, with MAE reductions of $\SI{12.5}{\%}$, $\SI{11.0}{\%}$, and $\SI{9.0}{\%}$, respectively. At these time-slot indices, only a limited number of RTT observations collected while connected to the current serving satellite are available. RT-LSTM compensates for this limited RTT history by directly leveraging real-time serving-satellite and gateway-relative features. As more time elapses after the scheduling instant, additional RTT observations become available to the baseline LSTM, allowing it to capture the current RTT pattern more effectively and thereby narrowing the performance gap.

\subsubsection{Performance variation across reference time-slot indices and prediction steps}
As shown in Fig.~\ref{fig:mae_combined}(a), the absolute MAE generally increases as the reference time advances through the scheduling cycle toward the next scheduling instant, because a larger portion of the $\num{15}$-step prediction horizon extends beyond the next scheduling instant. Since the serving satellite after the next scheduling instant is unknown at the reference time, its future features cannot be incorporated into the prediction. For the same reason, Fig.~\ref{fig:mae_combined}(b) shows a sharp increase in MAE at $t+15$ at reference time-slot index $\num{12}$ (the scheduling instant), because this prediction step coincides with the next scheduling instant and the serving-satellite information thereafter is unavailable. A similar increase is observed at the prediction step that crosses the next scheduling instant for the other reference time-slot indices.

\subsubsection{Gateway-relative features}
Gateway-relative features derived from Japanese gateways exhibit strong correlations with RTT. In particular, the satellite azimuth angle relative to the Otaru gateway and the satellite distance to the Yamaguchi gateway rank first and second in correlation magnitude, with coefficients of $\num{0.264}$ and $\num{0.257}$, respectively. These results indicate that gateway-relative geometry is closely associated with RTT variation across the link modes.

\FloatBarrier
\section{Conclusion}
In this paper, we addressed the blind-zone limitations of existing scheduling-cycle-based serving-satellite identification methods by developing StarRTT, an end-to-end real-time framework for Starlink LEO networks. Through comprehensive empirical analysis across multiple satellite generations (V1.0, V1.5, and V2.0-mini), we identified a bent-pipe coverage area and observed distinct RTT characteristics between bent-pipe and ISL modes. Based on these observations, we proposed RT-LSTM within this real-time framework, which jointly leverages historical RTT observations, real-time serving-satellite and gateway-relative features, and TLE-propagated future features through historical and future LSTM branches with gated fusion. Under strict real-time back-testing, RT-LSTM achieved an MAE of \SI{1.405}{ms} averaged across all $\num{15}$ prediction horizons and $\num{15}$ reference time-slot indices, reducing the MAE by \SI{4.3}{\%}, \SI{11.6}{\%}, and \SI{22.1}{\%} compared with the existing baseline models, LSTM, StarNet, and ASTM, respectively.


\FloatBarrier

\end{document}